\documentclass[11pt,a4paper]{article}
\usepackage{amsmath,amssymb}
\usepackage{epsfig,graphicx}
\usepackage[english]{babel}
\usepackage{verbatim}
\usepackage{cite}

\def\starup#1{\mbox{$\raise1.8ex\hbox{$*$} \kern-.7em#1$}}
\def\krup#1{\mbox{$\raise1.8ex\hbox{$+$} \kern-1.0em#1$}}

\begin{document}

\title{\bf Scalar Leptoquark Contributions
into  $l_i\to l_j\gamma$ Processes}
\author{A.~V.~Povarov$^a$\footnote{{\bf e-mail}: povarov@uniyar.ac.ru},
A.~D.~Smirnov$^{a}$\footnote{{\bf e-mail}: asmirnov@uniyar.ac.ru }
\\
$^a$ {\small Division of Theoretical Physics, Department of Physics,}\\
{\small Yaroslavl State University, Sovietskaya 14,}\\
{\small 150000 Yaroslavl, Russia.}}

\date{}
\maketitle

\begin{abstract}
The contributions of scalar leptoquarks in lepton flavor violating (LFV) processes
of type $l_i\to l_j \gamma$ are investigated in frame of the minimal model with the four color symmetry
and Higgs mechanism of quark and lepton mass generation.
It is shown that experimental data on the decays
$\mu\to e \gamma$, $\tau \to \mu \gamma$, $\tau\to e \gamma$
allow the existence of light scalar leptoquarks of type under consideration, with masses of order 1 TeV
or below.
\end{abstract}

One of the possible variant of new physics beyond the SM can be the
variant induced by the four color symmetry between quarks and leptons
of Pati-Salam type \cite{PS}.
The  minimal realization this symmetry MQLS model \cite{AD1} is predicted the existence doublets scalar leptoquarks,
which appear to be some kind of partner of the standard Higgs doublet.
Current  limits on masses from direct search of scalar leptoquarks are small $M_{LQ} \sim 200-300$ GeV\cite{PDG},
 Indirect  limits from $K_L^0\to \mu^\pm e^\mp$\cite{AD2009}, $S,T,U$ parameters\cite{PovSm4},
   $g-2$ \cite{Pov06} and others are close to direct limits.
   Other source limit on the masses scalar leptoquark can be LFV processes.
  Exist strong experimental  limits on LFV processes
$$ Br(\mu\to e\gamma)< 1.2\cdot 10^{-11} \qquad  \mbox{\cite{Ahmed:2002}},$$
 $$Br(\tau\to \mu\gamma)< 4.5\cdot 10^{-8} \qquad \mbox{\cite{Hayasaka}},$$
  $$Br(\tau\to e\gamma)< 3.3\cdot 10^{-8} \qquad \mbox{\cite{Guido:2010yn}}.$$

The topic of my talk is investigated the contributions new physics
into processes with lepton flavor violation
  in framework the minimal four color
symmetry model.

 MQLS model
is based on the group
$$G=SU_V(4)\times SU_L(2)\times U_R(1).$$
In the MQLS model the basic left- (L) and right- (R) handed quarks
 $Q'^{L,R}_{i a \alpha}$
and leptons
$l'^{L,R}_{i a} $ form the fundamental quartets of $SU(4)$ color group,
and can be written, in general, as superpositions
of the quark and lepton  mass
eigenstates $Q^{L,R}_{i a \alpha}$ and $l^{L,R}_{i a}$
\begin{eqnarray}
\psi^{L,R}_{i a \alpha}=
{Q'}^{L,R}_{i a \alpha} = \sum_j \left ( A^{L,R}_{Q_a} \right )_{i j}
Q^{L,R}_{j a \alpha} , \,\,  
\psi^{L,R}_{i a 4}=
{l'}^{L,R}_{i a} = \sum_j \left ( A^{L,R}_{l_a} \right )_{i j}
l^{L,R}_{j a},
\nonumber     \label{eq:ql}
\end{eqnarray}
where i=1,2,3 are the generation indices $ a=1,2$ are the $SU_L(2)$ indices
and $A=\alpha,4$- $SU_V(4)$ indices $\alpha=1,2,3$ are the $SU_c(3)$ color
indices.
The unitary matrices $A^{L,R}_{Q_a}$ and $A^{L,R}_{l_a}$ describe the fermion
mixing and diagonalize the mass matrices of quarks and leptons.
This matrices combined $C_Q = (A^L_{Q_1})^+ A^L_{Q_2}$
the Cabibbo-Kobayashi-Maskawa matrix,
which is know to be due to the distinction between the mixing matrices
 $A^{L}_{Q_1}$ and  $A^{L}_{Q_2}$
in(\ref{eq:ql})
for up and down left-handed quarks,
 $C_l = (A^L_{l_1})^+  A^L_{l_2}$  the matrix that is analog
its in the lepton sector and
is not diagonal, this evident from neutrino oscillation,
which is due to the possible distinction between the mixing matrices
 $A^{L}_{l_1}$ and  $A^{L}_{l_2}$
and
$K^{L,R}_a = (A^{L,R}_{Q_a})^+ A^{L,R}_{l_a}$
 unitary  matrices additional fermion mixing in model.
 (that are due to the possible distinctions
between the quarks and leptons mixing matrices
 $A^{L,R}_{Q_a}$ and  $A^{L,R}_{l_a}$).

In gauge sector model predicted of two vector leptoquarks
$V^{\pm}_{\alpha \mu} (\alpha=1,2,3)$ and of additional neutral
 $Z'$ boson.

The scalar sector of the model, contains four multiplets,
which transformed according to
 $ (4,1,1), \quad (1,2,1), \quad (15,2,1), \quad  (15,1,0)$
representations with respectively
 $\eta_1$, $\eta_2$, $\eta_3$, $\eta_4$ \hspace{20mm}-- VEV.

The MQLS model is based on the Higgs mechanism of splitting of the
quarks and leptons masses and predicted
 in addition to vector leptoquarks,
the existence  of
the doublets of scalar leptoquarks.
 In  this approach, the SM Higgs doublet
$\Phi^{(SM)}$ appearing to be superposition of the doublets $\Phi^{(2)}_a$
and $\Phi^{(3)}_{15,a}$ representation (1.2.1) and (15.2.1) with
VEV $\eta_2$ and $\eta_3$, corresponding.
and colorless doublets $\Phi^{(3)}_{15}$, that mixing with the doublet
 $\Phi^{(2)}_{a}$ from the representation
(1,2,1) give Higgs doublets SM and additional doublets $\Phi'$. Here
$\eta=\sqrt{\eta_2^2+\eta_3^2}$ is the SM VEV.

 In particular, the representation (15,2,1)
is kept two doublets SLQs
\begin{eqnarray}
(15.2.1)
\left ( \begin{array}{c} S_{1\alpha}^{(+)}  \\
         S_{2\alpha}^{(+)}\end{array} \right );
\left ( \begin{array}{c} S_{1\alpha}^{(-)}  \\
         S_{2\alpha}^{(-)}\end{array} \right ),
\nonumber
\end{eqnarray}

with electric charges of the component scalar doublets
\begin{eqnarray}
Q_{em}\,\,\,\,\,\,\,
\left ( \begin{array}{c} 5/3  \\
         2/3\end{array} \right );
\left ( \begin{array}{c} 1/3  \\
        - 2/3\end{array} \right ).
\nonumber
\end{eqnarray}

The scalar leptoquarks with electric charge 2/3 are superpositions three
physical scalar leptoquarks $S_1,S_2,S_3$ and Goldstone mode $S_0$,
\begin{eqnarray}
S_2^{(+)}&=&\sum_m C_m^{(+)}S_m, \
\starup {S_2^{(-)}}=\sum_m C_m^{(-)}S_m.
\nonumber
\end{eqnarray}
where $C^{(\pm)}_m$  are the  elements of the complex unitary matrices
the mixing of scalar leptoquarks of electric charge 2/3.

In the unitary gauge the physical leptoquarks fields are as follows:
two of up component doublets leptoquark $S_1^{(+)}$ and $S_1^{(-)}$
of electric charge 5/3 and -1/3,
respectively, and three scalar leptoquarks $S_m (m=1,2,3)$ of electric  charge
2/3.

The lagrangian interaction scalar leptoquarks with quarks and charge leptons can be written
in the following form \cite{PovSm1},
\begin{eqnarray}
L_{ulS^{(+)}_1} &=& \bar u_{i\alpha }  \Big [ ( h^L_+)_{ij}P_L +
(h^R_+)_{ij}P_R  \Big ] l_{j} S_{1\alpha}^{(+)} + {\rm h.c.},
\nonumber\\
L_{dl S_m} &=& \bar d_{i\alpha} \Big [ (h^L_{2m})_{ij}P_L+(h^R_{2m})_{ij}P_R
\Big ] l_{j} S_{m\alpha} +{\rm h.c.}
\nonumber
\end{eqnarray}
here,
$u,\,\, d,\,\, l,$ are, relatively,  up- and down- quarks,
charge leptons
of the generation $i$,
$P_{L,R}$ are the left and right
projection operators,
$h^{L,R}$ are the coupling constant matrices for generations.

The expression for coupling constant
because of Higgs origin are proportional to the
ratios of fermion masses to the SM VEV.
This ratios are quite small for first and second generation fermion
$m_u/\eta,m_d/\eta,m_s/\eta\sim 10^{-5}$
and    $m_c/\eta,m_b/\eta\sim 10^{-2}$, but the ratio of the
t-quark is not small $m_t/\eta\sim 0.7$.
Therefore general contributions into couplings constants give ratio of the
t-quark mass to SM VEV.
We neglected in coupling constant all
fermionic masses, except mass t-quark.

The dominant contribution in the coupling constant can be written as
\begin{eqnarray}
(h^L_+)_{3j} &=& \frac{\sqrt 6 \, m_t}{2 \eta \sin\beta} \,
(K_1^LC_l)_{3j} \,
\nonumber \\
(h^R_+)_{3j} &=& - \frac{\sqrt 6 \, m_b}{2 \eta \sin\beta} \, (C_Q)_{33}
( K^R_2 )_{3j}
\nonumber \\
(h^{L,R}_{2m})_{3j} &=& - \frac{\sqrt 6 \, m_b}{2 \eta \sin\beta}
(K_2^{L,R})_{3j} \, c_m^{(\mp)}.
\nonumber
\end{eqnarray}
Take notice, that SLQ  of MQLS model are like  LQ of 3-$d$ generation \cite{PPSMPLA}.

The contribution of scalar leptoquarks into the  $l_i\to l_j \gamma$ Processes:
is described by the two diagrams in figure 1. Contributions of $S_m$ and $b$-quark
in the amplitude $l_i\to l_j \gamma$  suppress
by  $m_b^2/m_t^2$ comparing with ones of
$S^{(+)}_1$ and $t$-quark.

\begin{figure}[h]
\centerline{
\epsfxsize=.80\textwidth
\epsffile{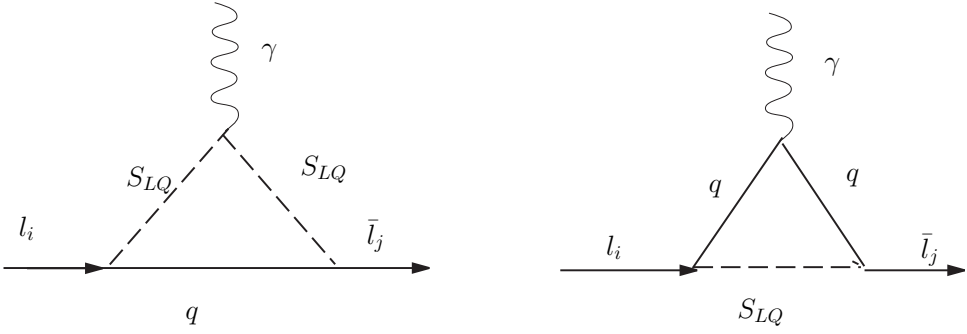}
}
\caption{Diagrams representing the contribution of scalar leptoquarks (SLQ) into the
 $l_i\to l_j \gamma$ Processes:  $q=u_i(d_i)$ is the up(down) quark of the $i$th
generation and
 $S_{LQ}=S_1^{(+)}(S_m)$ is the SLQ corresponding to the above quarks.}
\end{figure}\vspace{-2mm}

In general, the total one-loop contribution from this  diagrams
 can be written as
\begin{eqnarray}
 M &=& -\frac{ |e|}{64\pi^2m_{LQ}^2}\bar l_j \sigma^{\mu\nu}q^\nu
  \Bigg [
 m_i\bigg ( Q_kF_4(x) - Q_s F_2(x)  \bigg )
\bigg ( (|h^L|^2)_{ji}P_R +
(|h^R|^2)_{ji}P_L \bigg ) \nonumber\\ & + &
 2m_k\bigg ( Q_kF_3(x)  -
Q_s F_3(x)  \bigg )
\bigg ( (\krup{h^L}h^R)_{ji}P_R +
(\krup{h^R}h^L)_{ji}P_L   \bigg )  \Bigg ]l_i \epsilon^\mu ,
\nonumber
 \end{eqnarray}
 where  $Q_k$, $Q_S$ are electric charge of $q_k$-quark
and LQ, and  $x=m_k^2/m_{LQ}^2$,
\begin{eqnarray}
F_2(x)&=&\frac{1}{6(1-x)^4}(1-6x+3x^2+2x^3-6x^2\ln x), \nonumber \\
F_3(x)&=&\frac{1}{(1-x)^3}(1-x^2+2x\ln x), \nonumber \\
F_5(x)&=&\frac{1}{6(1-x)^4}(2+3x-6x^2+x^3-6x\ln x), \nonumber \\
F_6(x)&=&\frac{1}{(1-x)^3}(-3+4x-x^2-2\ln x). \nonumber
\end{eqnarray}
As can be seen from amplitude  the first term is proportional
to the initial lepton mass,
( whereas  the second term is proportional
to the quark mass.)

The probability this processes can be written as
\begin{eqnarray}
 W(l_i \to l_j\gamma)&=&
 \frac{9\alpha m_i }{256(4\pi)^4} (\frac{m_i}{\eta})^4 x^2
\bigg ( B_1^2(x) k^{(1)}_{ij} + 4(\frac{m_b}{m_i})^2 B_2^2(x) k^{(2)}_{ij}
\nonumber
\\ &-& 2\frac{m_b}{m_i}B_1(x)B_2(x)Re(k_{ij}^{(12)})
\bigg), \nonumber \\
B_1(x)&=&Q_k F_4(x)- Q_SF_2(x), \hspace{10mm}
B_2(x)=Q_kF_3(x)-Q_SF_1(x),   \nonumber \end{eqnarray}
here we proposes that there are t-quark in loop, and $k_{ij}$ is the matrices of mixing parameter in  the model
\begin{eqnarray}
 k^{(1)}_{ij}&=&\frac{|(K^L_1C_l)_{3j}|^2
|(K^L_1C_l)_{3i}|^2}{\sin^4\beta},\,\,\,\,  \nonumber\\
k^{(2)}_{ij}&=&\frac{1}{\sin^4\beta}\bigg (
|(K^L_1C_l)_{3j}|^2|(K_2^R)_{3i}|^2   + (i \leftrightarrow  j)
\bigg ),  \nonumber  \\
k^{(12)}_{ij}&=&\frac{1}{\sin^4\beta}\bigg (
(|(K^L_1C_l)_{3j}|^2 +  |(K_2^R)_{3j}|^2)\times \nonumber \\
&\times & ((K^L_1C_l)^*_{3i}(K_2^R)_{3i}  +
(K_2^R)^*_{3i}(K^L_1C_l)_{3i})
+ (i \leftrightarrow  j)    \bigg ).\nonumber
\end{eqnarray}
In probability decay second term are proportional
ratios of b-quark mass to the initial lepton mass (therefore this term give large contribution in numerical calculations).

Substituting the numerical values into eqnarray we obtain  following
expressions
\begin{eqnarray}
 Br(\mu\to e\gamma)&=&      1.1 \times 10^{-4} x^2 \bigg ( B_1^2(x) k^{(1)}_{\mu e} +
 \nonumber\\ &+&7056 B_2^2(x) k^{(2)}_{\mu e} -
 84B_1(x)B_2(x)Re(k_{\mu e}^{(12)})
  \bigg),\label{eq:n1}\\
 Br(\tau \to  e\gamma) &= &     2.2\times 10^{-5}
x^2 \bigg ( B_1^2(x) k^{(1)}_{\tau e} +\nonumber\\
&+& 20 B_2^2(x) k^{(2)}_{\tau e} -
4.8B_1(x)B_2(x)Re(k_{\tau e}^{(12)})
 \bigg), \nonumber\\
 Br(\tau \to  \mu\gamma)&= &     2.2\times 10^{-5}
x^2 \bigg ( B_1^2(x) k^{(1)}_{\tau \mu} + \nonumber\\
&+& 20 B_2^2(x) k^{(2)}_{\tau \mu} -
4.8B_1(x)B_2(x)Re(k_{\tau \mu}^{(12)})
 \bigg). \nonumber
\end{eqnarray}
By virtue of the fact that experimental restrictions on the $\tau$ -- decays are weaker than on the
$\mu$ -- decay, as results of the corresponding limitation on  the masses SLQ are small. Thus
we examine in the beginning $\mu$ -- decay.

{\bf I)} The General Contribution: we  retain only the second
term  in eqnarray (\ref{eq:n1}) in our numerical calculations, then this expression can be written as
\begin{eqnarray}
Br(\mu\to e\gamma) =0.7 x^2B_2^2(x)k^{(2)}_{\mu e},
\nonumber
\end{eqnarray}
\vspace{-10mm}
\begin{eqnarray}
k^{(2)}_{\mu e}=\frac{1}{\sin^4\beta}\bigg (
|(K^L_1C_l)_{3e}|^2|(K_2^R)_{3\mu}|^2   + (\mu \leftrightarrow  e)
\bigg ).  \nonumber
\end{eqnarray}
The expression are simple for analysis.
The lower limit on masses scalar leptoquark
 $S_1^{(+)}$ from decay
  $\mu\to e\gamma$  for different value parameter
  $k_{\mu e}^{(2)}$ shown on picture 2.
\begin{figure}[h]
\centerline{
\epsfxsize=.75\textwidth
\epsffile{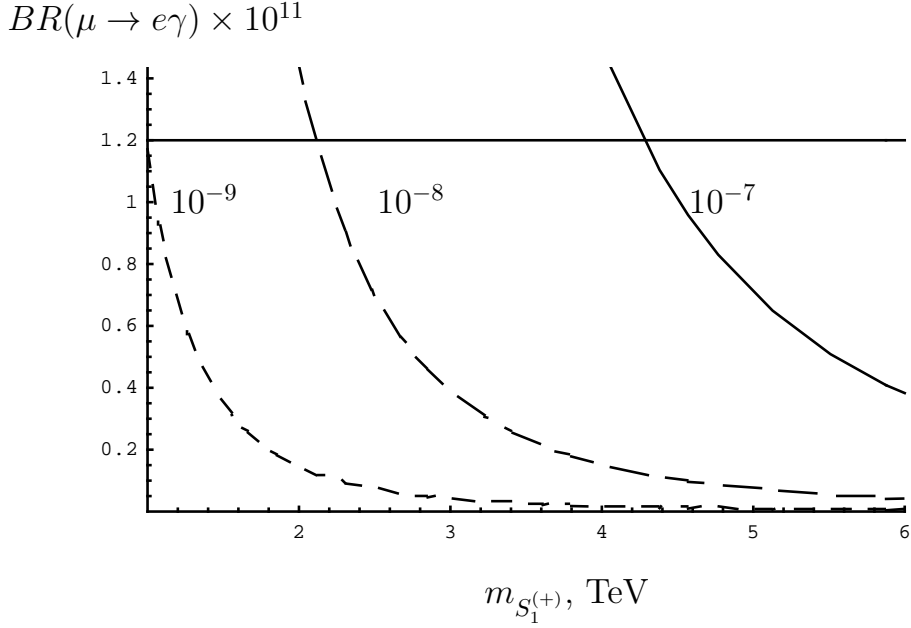}
}\caption{The lower limit on mass scalar leptoquark
  $S_1^{(+)}$ resulting from
  $BR(\mu\to e\gamma)$ with  value of parameter
  $k_{\mu e}^{(2)}=10^{-9},10^{-8},10^{-7}$ (Horizontal line corresponds experimental limit   $Br(\mu\to e\gamma)< 1.2\cdot 10^{-11}$)}
\end{figure}
As show mass scalar leptoquark
 $S_1^{(+)}$ can be below 1 TeV if matrix element $(K^R_2)_{13},(K^L_1C_l)_{13}
\sim 10^{-3}$ and $(K^R_2)_{23},(K^L_1C_l)_{23}\sim 10^{-2}$.

{\bf II)} In particular case, matrix  $K^R_2=I$,
 we  retain only the first term in eqnarray (\ref{eq:n1}):
\begin{eqnarray}
Br(\mu\to e\gamma) =1.1\times 10^{-4} x^2B_1^2(x)k^{(1)}_{\mu e},
\nonumber\end{eqnarray}
\vspace*{-7mm}
\begin{eqnarray}
  k^{(1)}_{\mu e}&=&\frac{|(K^L_1C_l)_{3e}|^2
|(K^L_1C_l)_{3\mu}|^2}{\sin^4\beta}.
\nonumber\end{eqnarray}

\begin{figure}[htbp]
\centerline{
\epsfxsize=.75\textwidth
\epsffile{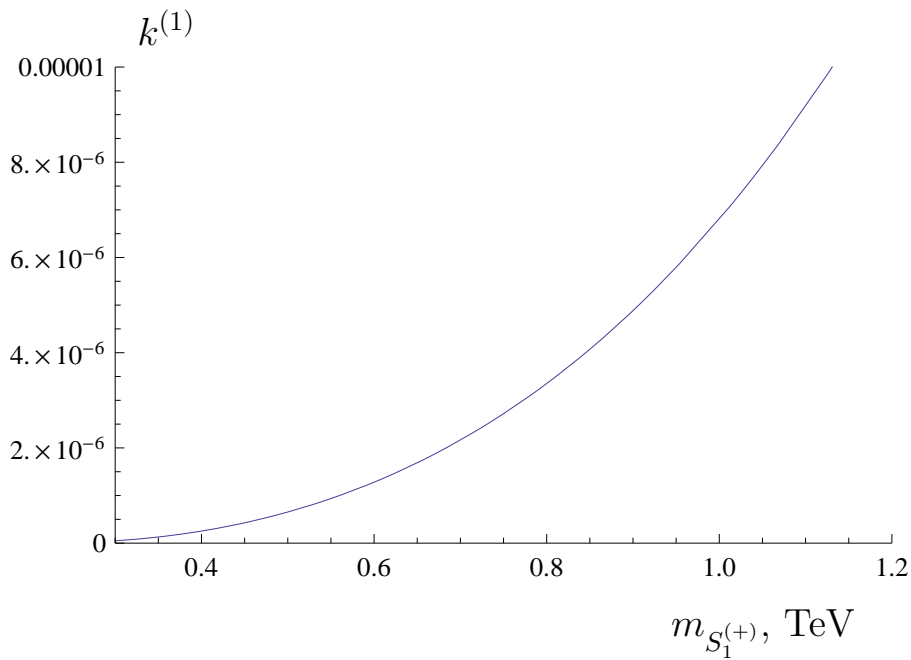}
}\caption{The allowed region of parameters
$m_{S_1^{(+)}}$ and  $k_{\mu e}^{(1)}$  resulting from $BR(\mu\to e\gamma)$ with  $K^R=I$( below the curve)  .}
\end{figure}
In  figure 3 shown, the region possible parameters $m_{S^{(+)}_1}$ and $k_{\mu e}^{(1)}$
resulting from  $BR(\mu\to e\gamma)$ with $K^R_2=I$.
As show masses LQ can be below 1 TeV if matrix element
  $(K^L_1C_l)_{13}
\sim 10^{-2},$  $(K^L_1C_l)_{23}\sim 0.1$,
 this restriction is weak that in previous Variant.

{\bf IIa)} In particular case  $K_2^L=K_2^R=I$  mixing parameter  $k_{\mu e}^{(1)}$ can be written in the  form
\begin{eqnarray}
  k^{(1)}_{\mu e}&=&\frac{|(U)_{13}|^2
|(U)_{23}|^2}{\sin^4\beta}.
\nonumber\end{eqnarray}
where
$\krup{C_l}=U_{PMNS}\equiv U$
 is Pontecorvo-Maki-Nakagawa-Sakata matrix and $U_{13}$ is its unknown element.

 \begin{table}[h]
\center{
\begin{tabular}{|l|l|l|l|}\hline\rule[-2ex]{0ex}{5ex}
$m_{S_1^{(+)}}$\,\,\mbox{TeV} &\,\,\,$0.55$ &\,\,\,$1.3$
&\,\,\, $9.3$
\rule[-2ex]{0ex}{5ex}\\ \hline
 $\sin\beta=0.2$\,  &\,\, $6\times 10^{-5}$ &\,\,\,
$2\times 10^{-4}$&\,\,\,$6\times 10^{-3}$
  \rule[-2ex]{0ex}{5ex}\\ \hline
 $\sin\beta=1$\,  &\,\, $1\times 10^{-3}$ &\,\,\,
$5\times 10^{-3}$&\,\,\,$0.14$
  \rule[-2ex]{0ex}{5ex}\\ \hline
\end{tabular}
\caption{Upper limit on the matrix element $U_{13}$
resulting from process  $\mu\to e\gamma$ with
 $K^{R,L}=I$ in dependence on
 $\sin\beta$ and  $S_1^{(+)}$ mass.
}}
\end{table}\vspace{-3mm}
In table 1  for example given up limit value matrix element $U_{13}$
for different mass scalar leptoquarks and model parameter $\sin\beta=\eta_3/\eta$.
  As can be seen value in model lower than experimental restriction $U_{13}<0.16$
In figure 4 shown the up limit matrix element $|U_{13}|^2$ as function
 mass scalar leptoquark. The current experimental limit are  $|U_{13}|^2< 0.032$\cite{Fogli06}.
 \begin{figure}[h]
\center{
\epsfxsize=.85\textwidth
\epsffile{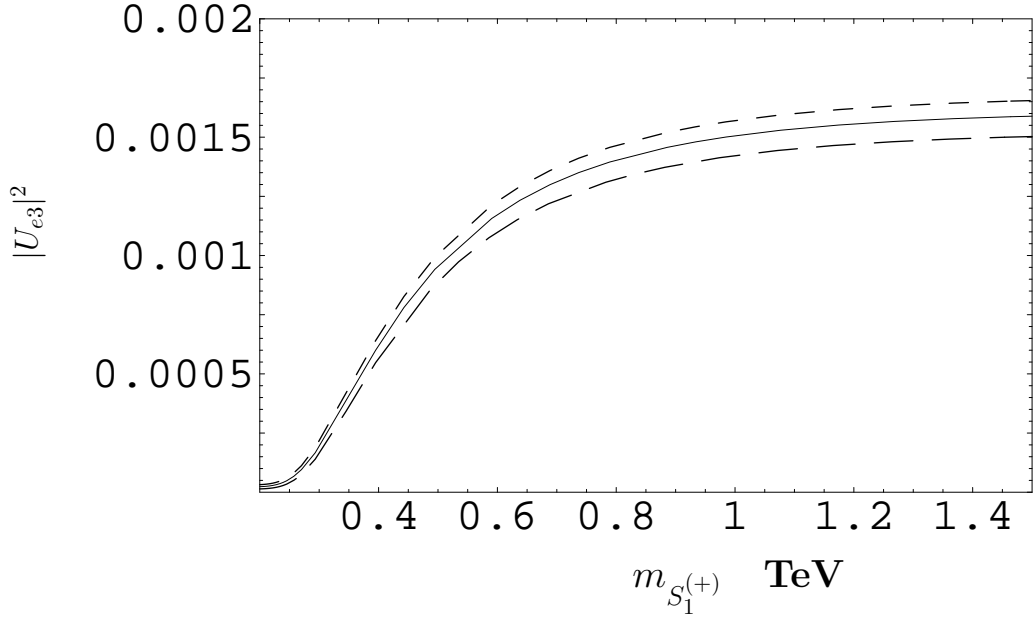}
\caption{Upper limit on the matrix element  $U_{13}$
resulting from  $\mu\to e\gamma$ with
 $K^{R,L}=I$ as  function of
 $S_1^{(+)}$ mass at
 $U_{23}=0.7^{+0.12}_{-0.16}$ \cite{Gonz} and $\sin\beta = 1$.
Current experimental limit
is  $|U_{13}|^2 < 0.032$ \cite{Fogli06} .}
}
\end{figure}
 As can be seen  scalar leptoquarks with arbitrary masses not restrict from $|U_{13}|^2$.

{\bf III)} Variant where  matrix elements $(K^L_1C_l)_{13}$,   $(K^R_2)_{13}$ are zero. When  only processes $\tau\to\mu\gamma$ exist and from  its experimental restriction $Br(\tau\to \mu\gamma)< 4.5\cdot 10^{-8}$ we obtain corresponding limits.
Lower limit on the  mass scalar leptoquark $S_1^{(+)}$  resulting
from  $\tau\to \mu\gamma $ in dependence on different parameter
 $k_{\tau\mu}^{(a)}, a=1,2,12$ shown in table 2.
 \begin{table}[h]
\center{
\begin{tabular}{|l|l|l|l|}\hline\rule[-2ex]{0ex}{5ex}
 $k_{\tau\mu}^{(1)}=k_{\tau\mu}^{(2)}=k_{\tau\mu}^{(12)}$
 &\,\, $10^{-4}$ &\,\,\,
$ 10^{-3}$&\,\,\,$ 10^{-2}$
  \rule[-2ex]{0ex}{5ex}\\ \hline
$m_{S_1^{(+)}}$\,\,\mbox{TeV} &\,\,\,$0.3$ &\,\,\,$0.7$
&\,\,\, $1.4$
\rule[-2ex]{0ex}{5ex}\\ \hline
\end{tabular}\caption{Lower Limit on the  $S_1^{(+)}$ Mass resulting
from  $\tau\to \mu\gamma $ in dependence on
 $k_{\tau\mu}^{(1)} = k_{\tau\mu}^{(2)} = k_{\tau\mu}^{(12)}$.
}}
\end{table}\vspace{-3mm}
 As shown this restriction is small.

{\bf IV)} Interaction of scalar leptoquark
$S_m$\,\,\, $(Q=2/3)$ and $b$-quark
gives the lower limit on $m_{S_m}$ from  $\mu\to e\gamma$
weaker than the current experimental ones.

{\bf  V)}  The Case of the chiral interaction of  scalar leptoquarks  $S_1^{(+)}$ with fermions gives the limits
which coincide with those of the
  [Variant II].

\vspace{3mm}
{\bf  Conclusion}

The contributions of scalar leptoquarks  $S^{(+)}_1$, $S_m$
from the MQLS model in  $l_i\to l_j \gamma$ decays are analyzed
in comparison  with experimental data on
$\mu\to e \gamma,\,\,\tau\to \mu \gamma,\,\,\tau\to e \gamma$
decays.

It is shown that in the appropriate region of the mixing
parameters relatively light scalar leptoquarks
(with masses of order  1 TeV  or below)
do not contradict current experimental restrictions
on LFV processes.

\vspace{3mm}

{\bf Acknowledgements}

This work was supported in part by the Ministry of Education
and Science of the Russian Federation
 in the framework of realization of the Federal
Target Program
``Scientific and Pedagogic Personnel of the Innovation Russia'' for 2009 -
2013 (project no. P-795).

\end{document}